\documentclass[conference]{IEEEtran}
\IEEEoverridecommandlockouts
\usepackage{cite}
\usepackage{amsmath,amssymb,amsfonts}
\usepackage{algorithmic}
\usepackage{graphicx}
\usepackage{textcomp}
\usepackage{xcolor}
\usepackage{makecell}
\usepackage{lscape}
\usepackage{listings}
\usepackage{hhline}

\usepackage{url}
\def\BibTeX{{\rm B\kern-.05em{\sc i\kern-.025em b}\kern-.08em
    T\kern-.1667em\lower.7ex\hbox{E}\kern-.125emX}}
\begin{document}

\title{Utilising urgent computing to tackle the spread of mosquito-borne diseases}

\author{\IEEEauthorblockN{Nick Brown}
\IEEEauthorblockA{\textit{EPCC} \\\textit{The University of Edinburgh}\\
Edinburgh, UK \\
n.brown@epcc.ed.ac.uk}

\and
\IEEEauthorblockN{Rupert Nash}
\IEEEauthorblockA{\textit{EPCC} \\\textit{The University of Edinburgh}\\
Edinburgh, UK}

\and
\IEEEauthorblockN{Piero Poletti}
\IEEEauthorblockA{\textit{Bruno Kessler Foundation}\\
Trento, Italy}

\and
\IEEEauthorblockN{Giorgio Guzzetta}
\IEEEauthorblockA{\textit{Bruno Kessler Foundation}\\
Trento, Italy}

\and
\IEEEauthorblockN{Mattia Manica}
\IEEEauthorblockA{\textit{Bruno Kessler Foundation}\\
Trento, Italy}

\and
\IEEEauthorblockN{Agnese Zardini}
\IEEEauthorblockA{\textit{Bruno Kessler Foundation}\\
Trento, Italy}

\and
\IEEEauthorblockN{Markus Flatken}
\IEEEauthorblockA{
\textit{German Aerospace Center (DLR)}\\
Braunschweig, Germany}

\and
\IEEEauthorblockN{Jules Vidal}
\IEEEauthorblockA{
\textit{Sorbonne Université}\\
Paris, France}

\and
\IEEEauthorblockN{Charles Gueunet}
\IEEEauthorblockA{
\textit{Kitware}\\
Lyon, France}

\and
\IEEEauthorblockN{Evgenij Belikov}
\IEEEauthorblockA{\textit{EPCC} \\\textit{The University of Edinburgh}\\
Edinburgh, UK}

\and
\IEEEauthorblockN{Julien Tierny}
\IEEEauthorblockA{
\textit{Sorbonne Université}\\
Paris, France}

\and
\IEEEauthorblockN{Artur Podobas}
\IEEEauthorblockA{\textit{KTH Royal Institute of Technology}\\
Stockholm, Sweden}

\and
\IEEEauthorblockN{Wei Der Chien}
\IEEEauthorblockA{\textit{KTH Royal Institute of Technology}\\
Stockholm, Sweden}

\and
\IEEEauthorblockN{Stefano Markidis}
\IEEEauthorblockA{\textit{KTH Royal Institute of Technology}\\
Stockholm, Sweden}

\and
\IEEEauthorblockN{Andreas Gerndt}
\textit{German Aerospace Center (DLR)}\\
Braunschweig, Germany}

\maketitle

\begin{abstract}
It is estimated that around 80\% of the world's population live in areas susceptible to at-least one major vector borne disease, and approximately 20\% of global communicable diseases are spread by mosquitoes. Furthermore, the outbreaks of such diseases are becoming more common and widespread, with much of this driven in recent years by socio-demographic and climatic factors. These trends are causing significant worry to global health organisations, including the CDC and WHO, and-so an important question is the role that technology can play in addressing them.
In this work we describe the integration of an epidemiology model, which simulates the spread of mosquito-borne diseases, with the VESTEC urgent computing ecosystem. The intention of this work is to empower human health professionals to exploit this model and more easily explore the progression of mosquito-borne diseases. Traditionally in the domain of the few research scientists, by leveraging state of the art visualisation and analytics techniques, all supported by running the computational workloads on HPC machines in a seamless fashion, we demonstrate the significant advantages that such an integration can provide. Furthermore we demonstrate the benefits of using an ecosystem such as VESTEC, which provides a framework for urgent computing, in supporting the easy adoption of these technologies by the epidemiologists and disaster response professionals more widely.
\end{abstract}


\section{Introduction}

The spread of diseases driven by mosquitoes as carriers, or vectors, between an infected individual and the as-yet uninfected healthy population accounts for around 20\% of the global burden of communicable diseases \cite{who}. The two major outbreaks of Chikungunya virus in Italy in 2007 and 2017, with the Aedes Albopictus mosquito as the vector, demonstrate that whilst one naturally associates such challenges with tropical climates, mosquito-borne diseases also effect other regions of the world including Europe and the US. The outbreak of Zika fever in 2016 in Brazil is another example where, by the end of the year it had spread, largely driven by mosquitoes, to over 48 countries.

The human infection that results from mosquito-borne diseases not only results in human health emergencies, but can have other knock on impacts too including the dwindling supply of blood, as donations have to be suspended due to infection risk. Currently there is significant concern in Europe because of the recolonisation of the Aedes Aegypti mosquito in Turkey, parts of the black sea and around the Mediterranean. This single insect is a vector for a wide variety of serious diseases, which include Dengue fever, Chikungunya virus, Zika fever, Mayaro and Yellow fever. Furthermore, the World Health Organisation (WHO), who have a specific agenda on global vector response, estimate that approximately 80\% of the world's population inhabit an area which is at risk of at-least one major vector borne disease. 

Consequently, by understanding in more detail the dynamics of mosquito abundance over time, one then possesses a crucial ingredient in assessing the risk of vector borne disease outbreaks, as well as being able to investigate their temporal patterns. To enable this we must generate highly accurate high dimensional output that estimates the risk of transmission for different diseases, which is able to be rapidly elaborated with updated input data, and whose complex and large outputs can be navigated with ease.

The main goal of this work is to integrate an epidemiological ensemble model with data analytics processes and visualization tools to support public health officials in making accurate data driven decisions. Such a tool can then be used to assess the risk associated with existing local outbreaks, identify those areas at increase risk of future outbreaks, and identify the optimal time for specific control measures to be applied. The rest of this paper is structured as follows; In Section \ref{sec:model} we briefly describe the intrinsic features of a mosquito-borne disease simulator that forms the foundation of this work before describing in Section \ref{sec:vestec} the VESTEC urgent computing ecosystem that we are looking to integrate with. We then explore details of this integration in Section \ref{sec:vestec_integration}, before illustrating how the human health professionals interact with the system and explore results in Section \ref{sec:health_interaction}. We then discuss the trickling of simulation results back to the user by running a mixture of shorter coarse-grained simulations and longer more accurate fine grained configurations in Section \ref{sec:optimise}, before drawing conclusions and describing further work in Section \ref{sec:conclusions}.

\section{Modelling of mosquito-borne diseases}
\label{sec:model}
The risk associated with vector borne diseases largely depends on the abundance of competent vectors, the density of mosquitoes in our case, for transmission of the infection. The Centre for Disease Control (CDC) Epidemic Prediction Initiative stresses that estimates of the potential vector density is a crucial issue for the public health decision makers. However traditional computational models are not able to estimate the absolute abundance of different mosquitoes species at high resolution over large spatial scales, nor to update these estimates in near real-time for reflecting changes in the meteorological or epidemiological conditions. 

Risks associated with mosquito-borne diseases are computed by considering a model which, 
when calibrated with a large set of entomological data collected on adult mosquito captures, enables high resolution estimates of mosquito populations across large regions. This approach extends a previously published method used to model the spread of Zika fever across the Americas in 2016 \cite{zhang2017spread}. The accuracy delivered by this model is based on two factors, firstly that the overall abundance of the mosquitoes is driven by a variety of socio-demographic and climatic factors, including gross domestic product (GDP), human density, temperature, and precipitation. The second underlying idea of the model is that an increase in the abundance of adult mosquitoes occurs mainly because of persisting favourable temperature conditions over a certain period.

It is possible to work at the city, regional, country, or even continental scale when modelling the growth of a mosquito population, with the model operating at a spatial resolution of 250m regardless of the size of region of interest. The model works in two stages, firstly it is calibrated across different species of mosquito based upon the input temperature, precipitation records, human density, GDP, and detailed mosquito capture data. The last argument, the capture data, is expressed as the likelihood of observing a specific number of captured female adult mosquitoes over time, across different years and geographical locations. Once calibrated, the model is then able to produce estimates on the absolute abundance of selected mosquito species and the consequent risk of transmission for different diseases for each day of the year based upon varying the temperature, precipitation, human density, and GDP. 

The computational model is currently very specialist and requires significant knowledge on behalf of the research scientist to correctly configure the inputs and then analyse and interpret the resulting outputs. Instead, the objective is for this to be viewed far more as a \emph{black box}, where public health professionals who are likely not research science experts, are able to drive the simulation using a much more easy to understand high level view of the scenario and consume the data in a more abstracted form. Furthermore, the model is more computationally intensive than previous generation simulation approaches, meaning that current runs on workstations rather than HPC machines are limited to a smaller geographic area.


\section{VESTEC: An ecosystem for urgent computing}
\label{sec:vestec}
\begin{figure}[h]
\centering
\includegraphics[scale=0.08]{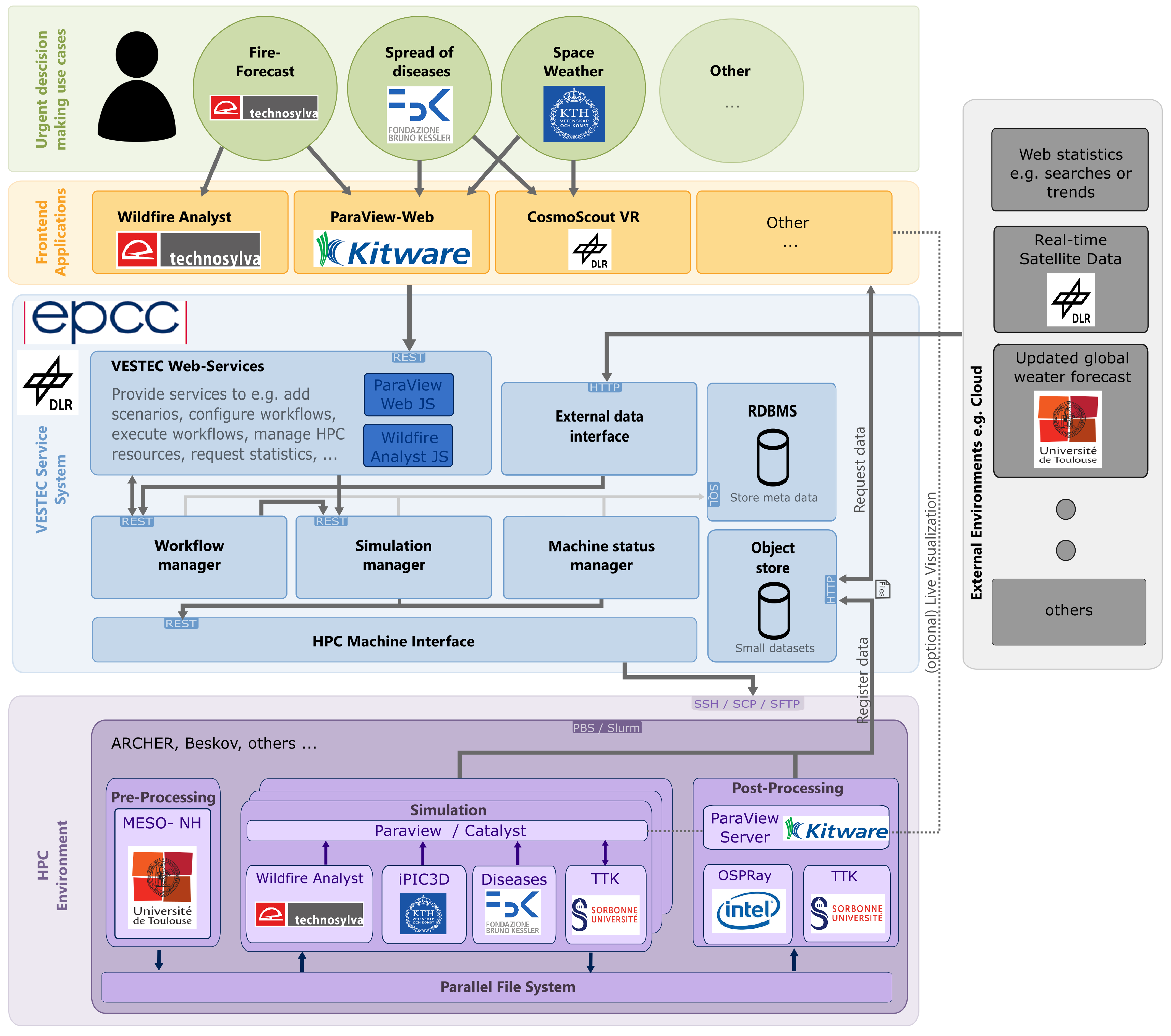}
\caption{Illustration of overall VESTEC system vision and constituent components}
\label{fig:vestec_overall}
\end{figure}

There were two major challenges identified with the epidemiological model in Section \ref{sec:model}; lowering the barrier of entry for public health professionals to configure the model and to consume the vast amount of output data, and the ability to accelerate the model via HPC. The Visual Exploration and Sampling Toolkit for Extreme Computing (VESTEC) system provides a technological solution to both these challenges, and Figure \ref{fig:vestec_overall} illustrates the overall vision of the VESTEC system which is designed to be a general purpose \emph{one stop shop} for supporting the execution of urgent workloads. This is illustrated in the top of Figure \ref{fig:vestec_overall} in green with a variety of urgent decision makers across numerous domains relying on the system. These users interact via a number of different platforms, from domain specific web-interfaces, to the popular ParaView \cite{ParaView} application, and CosmoScout VR \cite{COSMOSCOUT2021}, to drive disaster scenario simulations and consume results. Regardless of the specific user interface in use, they are able to present a mixture of real-time data reported from the field, such as temperature or precipitation, and the results of specific simulations scenario runs that the decision maker has undertaken. By using these interfaces the decision maker is manipulating the virtual environment, with HPC simulations then being executed transparently as required  to enable users to explore possibilities and the impacts of mitigation activities.


Whilst the computational power provided by supercomputers is key in enabling the timely simulations required for urgent workloads \cite{brown2019role}, such machines are often optimised for throughput rather than minimising the latency of individual jobs. This unbounded nature of the wait time in the queue means that in the past HPC machines are generally acknowledged as being less suited to running urgent workloads within specific time constraints. Whilst there has been previous work looking to address this, for instance SPRUCE \cite{beckman2007spruce} in the 2000s, much of this relied upon a manual approach of special access tokens that would provide queue priority. By contrast, the dynamic nature of disasters means that, to be driven most effectively by front-line responders in near real-time working interactively and driven by data arriving from the field, then exactly when simulations need to be run can be  highly unpredictable and could potentially involve numerous components coupled together. As disasters unfold such workloads can change dramatically, for instance driven by conditions in the field.

Consequently the VESTEC system follows the different approach of federating over multiple HPC machines. These machines can be physically located across a wide area, for instance the tier-0 supercomputers of Europe, where the VESTEC system will identify the most appropriate machine to run on given details of the simulation, queue status of the machine, and location of existing artefacts such as datasets. Based on this it aims to make intelligent decisions about \emph{what to run where} in order to generate results most quickly. Furthermore this provides resilience against the failure of a specific HPC machine, as such workloads are re-run elsewhere in such cases. All these concerns are entirely abstracted from the user and they never need to be aware of where their simulations are physically running or the specifics of how, just that they are generating reliable insights.

\subsection{Marshalling and control in the VESTEC system}
\label{sec:vestec_marshalling}
The blue box in Figure \ref{fig:vestec_overall} contains the marshalling and control functionality of the VESTEC system, which drive the execution of workloads across the HPC machine(s). Workflows are a fundamental aspect \cite{gibb2020bespoke}, which represent the different stages of progression through a disaster's lifetime. The stages comprising a workflow are triggered by some combination of external stimulus and/or preceding workflow stages. Written in Python and built upon the RabbitMQ AMQP messaging technology \cite{vinoski2006advanced}, methods representing workflow stages are decorated with an annotation and registered with the workflow manager against a corresponding queue name. When a message is pushed to this corresponding queue then the stage will be activated, with the received message provided as an argument to the method. To integrate a disaster scenario such as the simulation of mosquito-borne diseases with the VESTEC system, one must develop these workflow stages in Python. Stages can undertake a wide range of functionality including data transformation, preparation and submission of job to an HPC machine, and data clean up activities. At any point during execution, individual stages can send messages to corresponding queues which will activate other stages.

One way in which a workflow stage is initiated is by the arrival of data via the External Data Interface (EDI). When a disaster scenario is activated it will register a number of handlers in the EDI which can either work in pull mode (where the system will periodically poll a data source for new data), or push mode (which listens for data being sent to the EDI from some source). Regardless of the arrival method, this data is packaged as a message and sent to the corresponding workflow manager queue specified upon handler registration. The VESTEC marshalling and control system itself is not intended to run on a supercomputer, so whilst workflow stages can run concurrently across the cores, for data processing of any level of complexity it is preferred that this is performed via a job on the HPC machine rather than a stage of the scenario's workflow. 

\begin{figure}[h]
\centering
\includegraphics[scale=0.38]{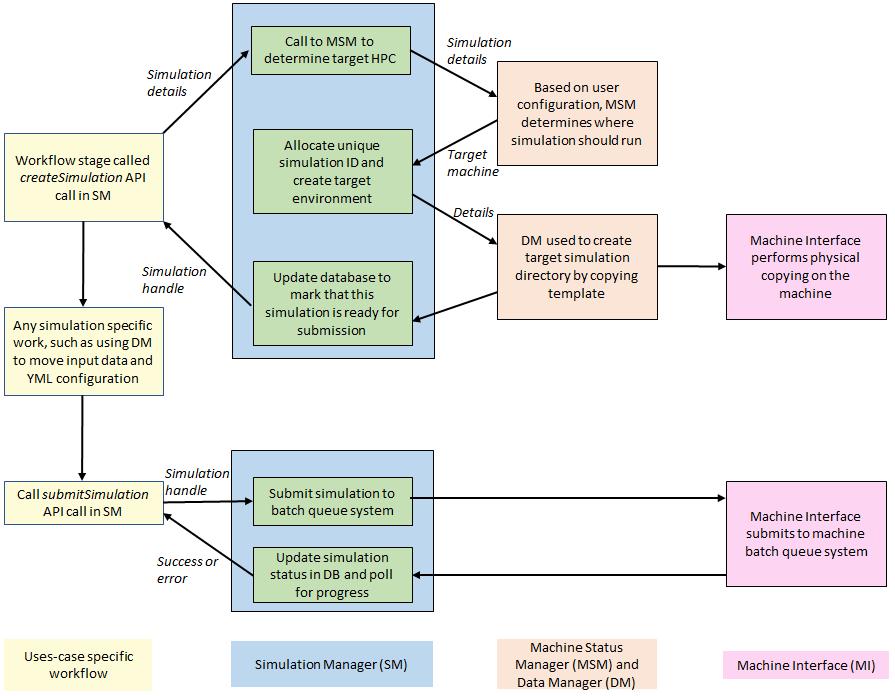}
\caption{Illustration of the interactions of VESTEC marshalling and control system stack involved in job submission on an HPC machine}
\label{fig:vestec_sim_submission}
\end{figure}

Much of the functionality in the blue box of Figure \ref{fig:vestec_overall} is infrastructure to support the requirements of the workflows. For instance, instead of communicating directly with the HPC machines, workflow developers use the well documented Simulation Manager (SM) API calls which enable them to prepare and submit jobs to the systems. There is a considerable amount of complexity in preparing and submitting a job, all of which is abstracted from the ultimate user and much from the workflow developer. This is illustrated in Figure \ref{fig:vestec_sim_submission}, where yellow boxes represent explicit activities in the workflow and the other boxes lower levels in the stack and their constituent activities deciding where a job should run and monitoring it. The job submission API accepts optional workflow stage callbacks, which will be activated when the job on the HPC machine reaches specific states, such as running, job completion, or an erroneous state.

The purple box of Figure \ref{fig:vestec_overall} represents the environment on the HPC machines, where numerous simulation codes and toolkits can run as part of a disaster scenario. To provide easy portability across HPC machines requires the ability to configure the individual simulations and couple them in a machine agnostic manner. To achieve this level of flexibility and generality the Common Workflow Language (CWL) \cite{crusoe2021methods} is used on the HPC machines. CWL is a mature standard for workflow description which has gained popularity in fields including bioinformatics. Consequently all workloads to be actually executed on the HPC machine are described in CWL and it is this workflow description that is submitted to the batch queue system, with CWL then deciding which applications to execute and when \cite{nash2020supercomputing}. 

In addition to providing a convenient approach for coupling application execution, CWL also solves the challenges around application configuration. This is because skeleton configurations can be provided via CWL, with runtime specific parameters then injected via YAML files. Such parameters can originate from the VESTEC system to configure each specific simulation, and furthermore additional sets can be provided on a machine by machine basis to specialise between supercomputers in an agnostic manner. Consequently, the core CWL configuration is machine agnostic, enabling quick and easy porting of simulation code configurations between supercomputers, with machine specific specialisation provided by a single point of truth YAML file.

From the discussion in this section it can be seen that workflows are a key enabler, not just for driving progression through the timeline of an incident, but also to enable convenient and machine independent configuration of simulations. The combination of RabbitMQ-based workflows in the VESTEC system, and CWL workflows on the HPC machines provide a choice around granularity. By encoding the entire execution on the VESTEC system side we maximise flexibility, but this results in overhead as there must be explicit queue submission for each simulation and communication between the VESTEC system and HPC machine. By contrast, the CWL workflows are atomic as far as the VESTEC system is concerned, with a series of coupled CWL workflow stages only requiring one queue submission and the completion of one CWL step automatically chaining to the next. The sweet spot around granularity depends on the disaster in question and target metrics.

\subsection{Visualisation}

CosmoScout VR \cite{COSMOSCOUT2021} is an open source modular virtual universe environment developed at the German Aerospace Center (DLR). Based upon virtual reality, it allows users to explore massive geo-referenced datasets and has been popular in fields including planetary science and climate change. Such capabilities are also especially applicable when navigating the population of mosquitoes. This immersive visualisation can be automatically augmented with facets including real-time observational data and results from simulations. CosmoScout VR has been integrated with the VESTEC system, and therefore from within this single virtual reality environment the urgent decision maker is able to explore their scenario and drive additional simulations based upon actions undertaken within the UI, with results then automatically presented when available. It is also possible to run CosmoScout VR in a more traditional two-dimensional mode, with results displayed on a monitor and using the GUI for interaction, and this is the form that we illustrate in Section \ref{sec:health_interaction}.

A key challenge associated with many computational simulations is handling the vast amount of output data which is generated. This is especially acute for disaster response, where urgent decision makers are often acting under severe time constraints but must still make the correct decision first time, every time. Therefore an important question is how the data can be post-processed to provide an accurate \emph{at-a-glance} view of the situation. Topological Data Analysis (TDA) \cite{carlsson2009topology} is one such solution, and this captures the features of interest in scalar data into concise, yet informative, topological data signatures. These signatures are typically orders of magnitude smaller than the data itself and can be used as a proxy to the data. There are numerous examples of the successful use of TDA across science and engineering, and an appealing aspect is the ease it offers for the translation of domain-specific feature descriptions into topological terms. There are numerous algorithms that are able to not only extract such topological features, but furthermore identify and eliminate unwanted aspects such as noise.  

The VESTEC system provides access to the Topology ToolKit (TTK) \cite{tierny2017topology} \cite{masood2019overview} which is an open source software library for TDA. Built upon the VTK/ParaView software ecosystem an analysis pipeline is developed using TTK modules in Python which executes these TTK modules for analysis and transformation. The result of running TTK is the generation of a series of persistence diagrams \cite{edelsbrunner2010computational} which visually represent the population of points of interest and their prominence. Available to run in both a post-processing mode on simulation output files, and also in-situ via the Catalyst coupler \cite{catalyst}, the output of this analysis pipeline can then be stored to disk in the form of a CINEMA database which is a SQL-type database of VTK files, or streamed over the network as appropriate. Another benefit of this approach is that because persistence diagrams, which are effectively topological signatures representing the original data, tend to be far smaller than the raw simulation data, it can reduce the overall data sizes that need to be transferred or stored. Whilst the use of topological proxies via TTK has significant advantages, both in terms of data presentation and data size storage and transfer, it does increase the overall amount of computation required. 

\section{Mosquito VESTEC integration}
\label{sec:vestec_integration}
\begin{figure}[h]
\centering
\includegraphics[scale=0.42]{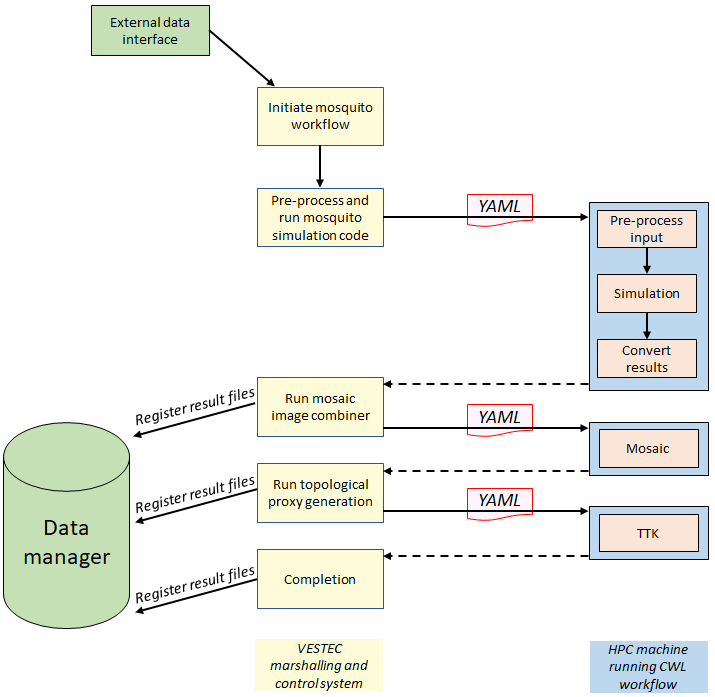}
\caption{Illustration of workflow developed for the mosquito use-case}
\label{fig:mosquito_workflow}
\end{figure}

Figure \ref{fig:mosquito_workflow} illustrates the workflow that has been developed to integrate support for mosquito-borne disease modelling into the VESTEC system. An incident is activated for a specific geographical region, mosquito type, and virus of interest, with this action of activation registering handlers in the EDI. These will listen for the arrival of data, either in the push or pull mode as described in Section \ref{sec:vestec_marshalling}, and activate the workflow initiation stage once data arrives. The initiation workflow stage will store such arrived data until all required input data-sets for a specific scenario are available and once all data is present the \emph{pre-process and run mosquito simulation code} stage is executed. The EDI and workflow is flexible enough such that this input data can be derived from different sources. For instance, CosmoScout VR can send through the entire inputs needed for temperature, precipitation, human density, and GDP, or alternatively a subset of these can originate from the GUI and the rest, especially temperature and precipitation, can be read from external sensors.

The \emph{pre-process and run mosquito simulation code} combines input data and generates the YAML configuration. Via the SM API calls (that were mentioned in Section \ref{sec:vestec_marshalling}), the appropriate HPC machine is identified and required inputs and configurations are transferred to it. Subsequently it will then instruct the SM to submit the simulation job to the HPC machine batch queue. As described in Section \ref{sec:vestec_marshalling}, this submission is a CWL script which will inject the run-specific YAML configuration into a template and execute the contained steps. In this instance there are three steps undertaken by the CWL workflow; pre-processing of the data (which would be too computationally intensive for the VESTEC system itself), execution of the mosquito simulation, and conversion of the simulation output into Tag Image File Format (TIFF) files. 

The subsequent VESTEC system workflow stage \emph{Run mosaic image combiner} is provided as a callback and will execute once the HPC machine simulation has run to completion. This will then schedule the mosaic post-processing of data, which itself will use the same callback mechanism to execute the \emph{Run topological proxy generation} VESTEC system workflow stage upon completion. This stage invokes the TTK processing of data on the HPC machine with the \emph{Completion} workflow stage executed upon completion. To optimise data size required for storage and transfer, a lossy topology-preserving compressor \cite{soler2018topologically} is used to compress the resulting TTK data files.

Figure \ref{fig:mosquito_workflow} also contains a \emph{Data Manager} (DM) component, with three of the workflow stages interacting with the DM to register data. The DM is responsible for tracking the location and status of files of interest, such as outputs that might be required by the user interface. Whilst it is possible for the DM to explicitly store files, this is discouraged to avoid excessive data overhead on the VESTEC marshalling and control system, and instead its main role is as a data directory. The DM contains services that will retrieve the data, streaming it back to the caller from its source location, such as the HPC machine, where it resides. For the mosquito integration there are three result data-sets of interest and these are stored in the DM as they become available. This is the reason for the design of the distribution of workflow stages between the VESTEC system and CWL because, as mentioned in Section \ref{sec:vestec_marshalling}, from the VESTEC system side the execution of a specific CWL workflow is atomic. Therefore by adopting this level of granularity the VESTEC system is activated when a code finishes which has generated a dataset of interest, registered in the DM, and able to make it available to the caller.

\section{Human health professional interaction}
\label{sec:health_interaction}
A module was developed for CosmoScout VR which enables the human health professional user to select mosquito scenario specific parameters such as temperature or human density, and present results returned from simulations in a time dependent fashion. They are able to overlay details over a geographical region, such as heatmaps of the inputs and R0 output and present these within the context of time. Furthermore, it is possible to set up specific filters in CosmoScout VR for these layers, meaning that the health professionals need only to focus on specific aspects, and also change the colours at specific thresholds which will better highlight features of interest. 

\begin{figure}[h]
\centering
\includegraphics[scale=0.33]{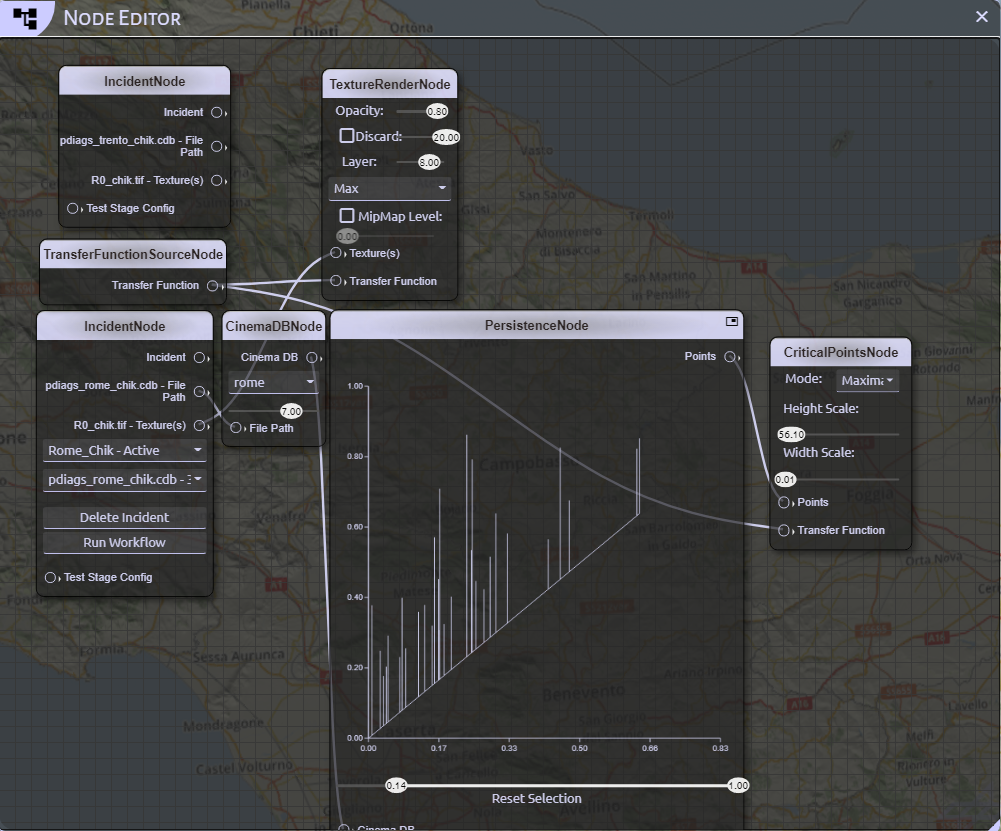}
\caption{Screenshot of CosmoScout VR node editor for mosquito-borne disease exploration}
\label{fig:node_editor}
\end{figure}

Figure \ref{fig:node_editor} illustrates the CosmoScout VR node editor, where a user has created and configured different nodes, which represent constituent components, and connected these together to drive how they are displayed. This configuration is exploring the spread of the Chikungunya virus via the Aedes Albopictus mosquito over Rome. Via the appropriate nodes the user can provide specific configuration parameters, which will then feed into the processing of preceding data and/or be sent to the simulations as they are executed.

\begin{figure}[h]
\centering
\includegraphics[scale=0.18]{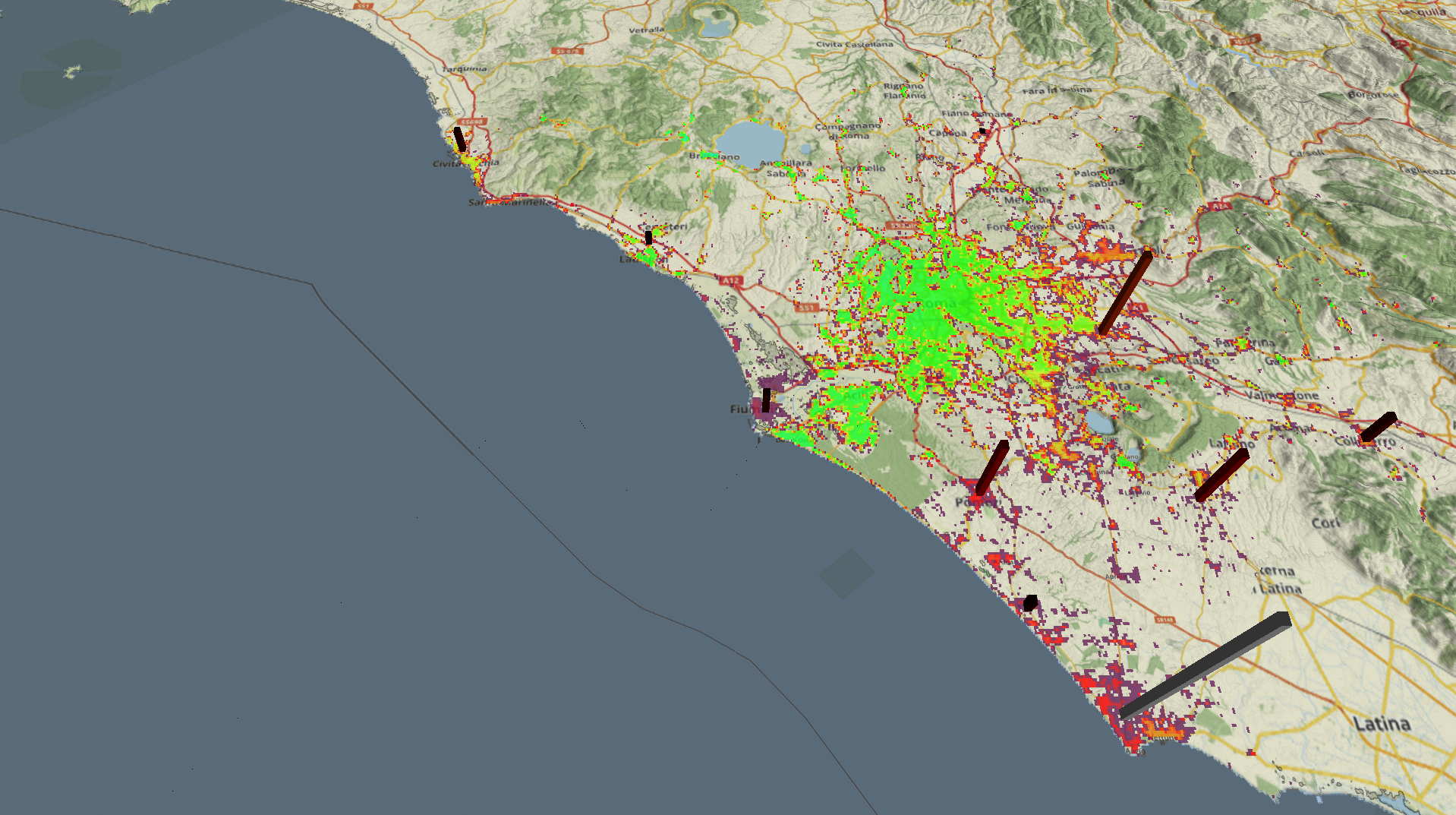}
\caption{Screenshot of CosmoScout VR overlaying the resulting R0 heatmap and persistence diagrams over Rome}
\label{fig:rome_ui_1}
\end{figure}

Figure \ref{fig:rome_ui_1} is a screen shot of the CosmoScout VR display pane which illustrates the overlaying of simulation results on-top of the map. This is for the same Chikungunya virus Rome scenario, and there are two overlays being displayed, firstly the R0 heatmap and secondly local maxima identified through the topological persistence diagrams. The R0 heatmap has been coloured from green to red, indicating areas from low potential disease spread in green to those in red with high risk. The vertical bars represent the local maxima and enable users to undertake a quick scan and identify those most critical and main infection risk areas.

The local maximima bars illustrated in Figure \ref{fig:rome_ui_1} are generated by TTK on the HPC machine and retrieved by CosmoScout VR when available. A brushing and linking approach was integrated with CosmoScout VR, where users can, via the node editor, brush data in the diagram and the selected data is then immediately highlighted in the view of CosmoScout VR. Due to the simulation resolution of 250m, the raw simulation output data that these diagrams are generated from is initially too sparse for use directly. Consequently, following the technique developed in \cite{vidal2021fast}, Gaussian resampling is undertaken on the sparse output field to generate a more fine grained piece wise linear scalar field to be used directly as an input to the topological proxy algorithms. Ultimately we generate a number of persistence diagrams, each covering a specific time period such as a day, week, or month, and the algorithms are configured to extract the extreme maxima of the field. This extreme maxima represents the areas with the highest R0 value, and depicts areas with highest risk to experience a spread of the disease. Furthermore an average of all the diagrams, known as the Barycentre, is calculated which can be used to characterise the entire time period of interest. This is useful as it enables convenient comparison between varying scenarios, for instance exploring the spread of disease with a different type of mosquito.

Whilst one might assume that identifying such maxima could be done via a simpler approach, the challenge is often that such maximum points in the raw data are typically located in the same geographical region. Instead, we want the maxima to be performed in a way that isolates regions of interest rather than just clustering within a specific region. This is where the algorithms provided by TTK are highly useful as they enable such identification and categorisation.

\begin{landscape}
\begin{figure*}
\hspace*{-7.5cm}
\centering
\includegraphics[scale=0.53]{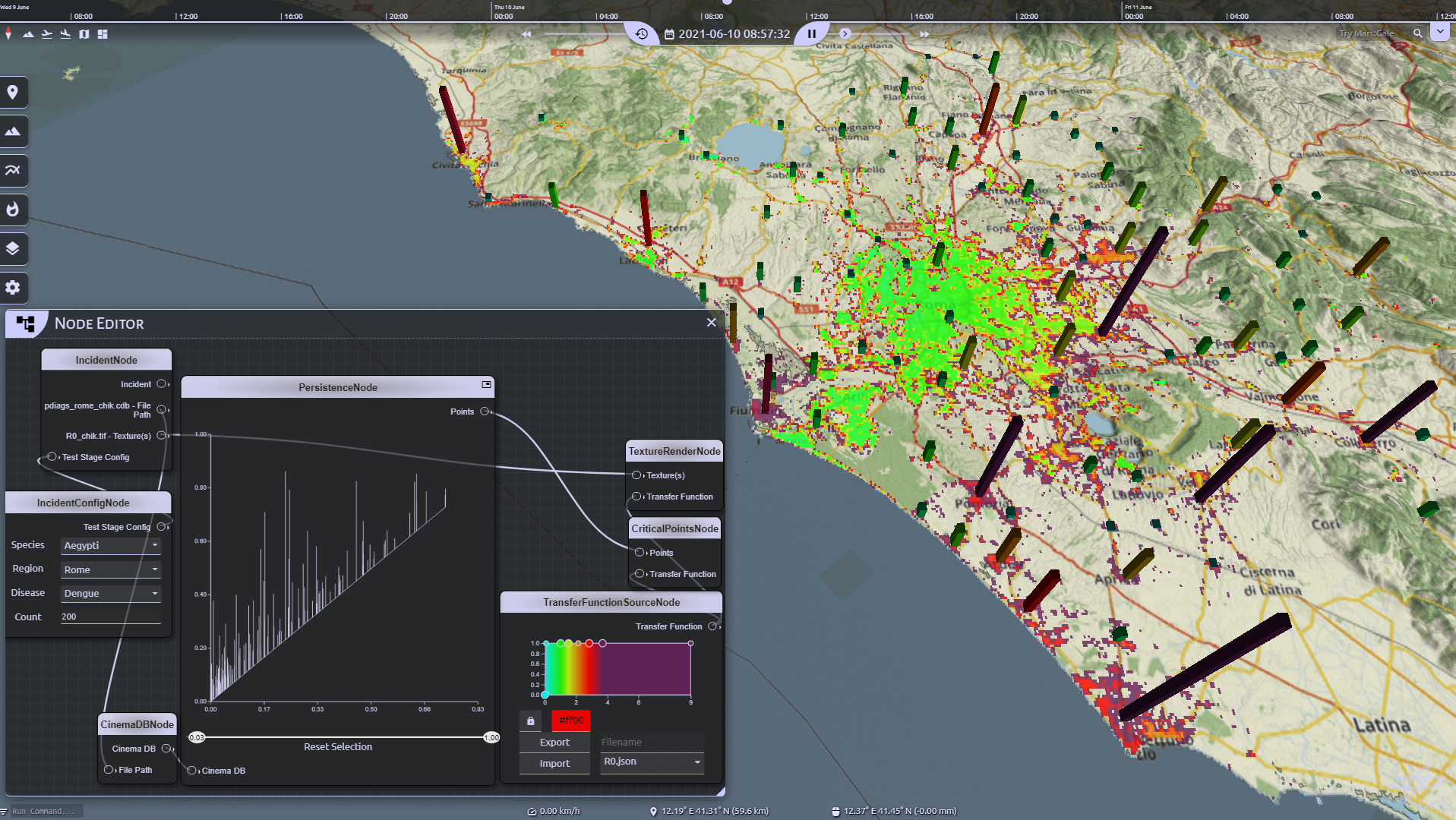}
\caption{Screenshot of CosmoScout VR GUI for spread of Chikungunya virus in Rome scenario}
\label{fig:full_ui}
\end{figure*}
\end{landscape}

Figure \ref{fig:full_ui} illustrates the full CosmoScout VR GUI for our mosquito integration, where the health professional is exploring the spread of Chikungunya virus via the Aedes Albopictus mosquito over Rome in more detail. Via the node editor it can be seen that they have applied filtering that considers a wider range of maxima R0 values for the persistence diagrams compared with the view of Figure \ref{fig:rome_ui_1}. Furthermore, using the controls at the top of the UI they are able to step through time, enabling users to explore how predicted changes to disease spread develop as the days and weeks progress. As the time of interest is changed, CosmoScout VR will retrieve the appropriate data files and extract the time points required. The coloured spectrum box contained within the node editor of Figure \ref{fig:full_ui} illustrates applying threshold colours to a field of interest, here the R0 heatmap, that will then be displayed to the user. Such configurations can be predefined, which is useful in a time critical situation because the health professional can quickly load known visualisation configuration constants for most accurately guidance.

\section{Trickling of partial simulation results}
\label{sec:optimise}
Health professionals exploring mosquito-borne disease spread via CosmoScout VR desire simulation results as soon as possible to aid them in making timely decisions. There are two aspects here, firstly the speed at which the simulation codes can run through to completion, and secondly the time queued up in the HPC machine batch system before execution begins on the supercomputer. As described in Section \ref{sec:vestec}, jobs running on HPC machines experience an unbounded queue time where there is no guarantee about the time frame that jobs will run in. 

Via the VESTEC system, the mosquito disease spread simulation code, data consumption, and workflows are flexible enough to support the submission of multiple applications for a specific scenario. Such submissions can range from coarse gained short running configurations to more detailed and longer executing versions. The idea is that the short running applications will generate results much more quickly, and although they might not be completely accurate or detailed, they provide a helpful preview while the more detailed configurations are still executing. Such coarse-grained results will then be replaced automatically by CosmoScout VR when the more accurate, longer running configuration results are available. The mosquito simulation code makes this convenient because it is running numerous ensembles, and by modifying this number we can control the granularity.

\begin{figure}[h]
\centering
\includegraphics[scale=0.55]{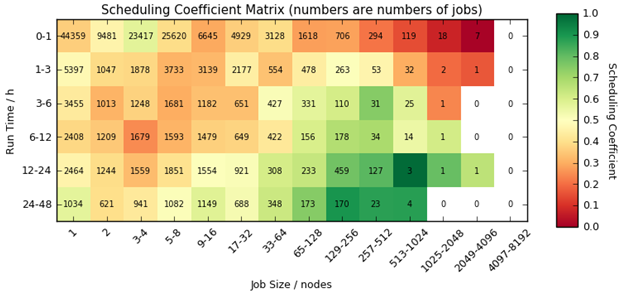}
\caption{Scheduling coefficient matrix for jobs running on ARCHER over a quarter (three months)}
\label{fig:sched_coeff}
\end{figure}

One might assume that shorter running jobs, requesting a smaller amount of parallelism and walltime, will progress faster through the batch queue than larger jobs. However this is a common but incorrect assumption. The behaviour of a typical batch queue is illustrated by Figure \ref{fig:sched_coeff}, which shows the scheduling coefficient matrix for ARCHER, the UK’s previous national supercomputer running between 2013 and 2020. This matrix represents all the jobs submitted within a quarter (over three calendar months), with the job size (in nodes) and runtime of the job (in hours) illustrated on the axis. For each box, which represents a specific configuration of job size and runtime, the included number is the number of jobs that corresponds to this configuration, and the colour represents the scheduling coefficient. This coefficient is the ratio of run time to run time plus queue time and, as such, represents the fraction of the job’s time in the queue that was spent running. A value of 1 indicates that the job started instantaneously and a value of 0.5 indicates that the job queued for the same amount of time that it ran for.

It can be seen from Figure \ref{fig:sched_coeff} that by far the most popular type jobs on ARCHER are those requesting small numbers of nodes and running over a short time frame. However, typically these do not start immediately, for example jobs requesting 1 node with up to an hour of runtime typically spend longer in the queue rather than the time that they run for. This means that if a user was to request a job of one node in size and this was to run for approximately one hour, then on average the job would have spent over an hour queuing before execution. Therefore, whilst it is likely that short jobs will spend a shorter absolute time in the queue than a larger job, this is by no means instantaneous and on average small, short jobs spend longer queuing then they do executing. 

However there is one situation where submitting short jobs to an HPC machine can reliably result in faster batch queue wait time, and that is where the machine provides a short queue. Short, or debug, queues typically enable small jobs, for instance those only running over a few nodes and running for a short amount of time (20 minutes maximum walltime being common), to be placed in a special high priority queue that will run quickly compared to the main submission queue. Consequently for specific HPC machines that provide this capability, if we are able to partition our simulation run into a coarse run that will fit into the short queue, and a longer run that will require the normal queue, then this can be advantageous. 

\begin{table}[h]

 \centering
\begin{tabular}{ | c c c c c c | }
\hline
\makecell{\textbf{Number} \\ \textbf{ensembles}} & \makecell{\textbf{Stage 1} \\ \textbf{time (sec)}} & \makecell{\textbf{Stage 2} \\ \textbf{time (sec)}} & \makecell{\textbf{Stage 3} \\ \textbf{time (sec)}} & \makecell{\textbf{Total} \\ \textbf{time (sec)}} & \textbf{Queue}\\\hline
10 & 16 & 4 & 66 & 86 & short \\
1000 & 309 & 4 & 85 & 398 & short \\
3000 & 1710 & 4 & 89 & 1803 & normal \\
\hline
\end{tabular}
\caption{Runtime of different mosquito-borne disease spread simulation configurations on Cirrus}
\label{fig:performance}
\end{table}

Table \ref{fig:performance} illustrates the runtime of different configurations of the mosquito-spread simulation over a node of Cirrus. Each Cirrus node contains two Intel Xeon E5-2695 (Broadwell) CPUs, each with 18 cores, and 256GB RAM. In Table \ref{fig:performance} we provide the runtime of each of the three jobs submitted to the HPC machine as described in Section \ref{sec:vestec_integration} and also the total time. It can be seen that 10 simulation ensembles involves a total of 86 seconds runtime, and 1000 ensembles 398 seconds. Both of these fit into the short queue, representing simulation runs that are able to deliver coarse-grained results after approximately a minute and a half, and then results of medium accuracy after around six and a half minutes. The most accurate simulation configuration, with 3000 ensemble members, requires 1803 seconds runtime, which is approximately thirty minutes, and this no longer fits into the short queue. Therefore for this more accurate configuration, in addition to the increased execution time it must also be submitted to the normal queue and will therefore likely experience a longer queue time which itself depends upon how busy the HPC machine is at the time of submission. 

\section{Conclusions and further work}
\label{sec:conclusions}
In this paper we have described the integration of an mosquito-borne disease spread model with urgent computing. Leveraging the VESTEC system, this has unlocked the convenient consumption and analysis of epidemiological data using HPC as the underlying computational powerhouse to run the required computational codes and undertake necessary data transformations. We have described in detail how the integration with the VESTEC system has been undertaken, and demonstrated the benefits that this ecosystem provides for making this technology available to human health professionals. 

Whilst it would be possible to manually integrate such a model with the HPC machines and constituent visualisation and analytics technologies, this would be time consuming and non-trivial. By contrast, the VESTEC system provides a ecosystem for urgent computing where, by the development of workflows and any bespoke UI components, one is able to leverage the underlying lower level functionality including HPC simulation marshalling and control, the tracking of data-sets and their transfer, the consumption of results, and appropriate data analytics to provide higher level quick to consume information.

Considering further work, we look to integrate in the future with a weather forecasting model. Currently it is possible to feed in real-time sensor data and for the user to be able to manipulate this via the CosmoScout VR user interface. However this provides limited opportunity to forecast the spread of these diseases in the coming days and weeks based upon how the weather might realistically change. In-fact other VESTEC urgent workloads, such as tracking the progression of forest fires, have been integrated with the high resolution Meso-NH model which undertakes high resolution modelling on latest Global Forecast System (GFS) data. As this weather model has already been successfully integrated with the VESTEC system it would likely be an obvious starting point, although this would require some data manipulation to extract the temperature and precipitation fields of interest as well as format these appropriately for the simulation codes. Furthermore we would like to address continent scale simulations of mosquito-borne disease spread, and this will require further optimisation at the individual disease spread modelling code level. Leveraging the existing visualisation and TTK analysis integrated in this work, many of the building blocks are already in place to that will enable us to present and organise the vast amounts of data that this increased scale will generate.

The result of this work is a significant advancement in human health capability, enabling these users to exploit urgent computing for predicting, analysing, and mitigating the spread of mosquito-borne diseases in near real-time. We have demonstrated the importance and role of urgent computing for these workloads, and the benefit of integrating with an existing ecosystem, such as VESTEC, which lowers the barrier to entry significantly. Consequently, this work also acts as a success story for the use of urgent computing and HPC machines in this regard, providing exciting new capabilities that deliver very significant societal benefits. 

\section*{Acknowledgment}

The research leading to these results has received funding from the Horizon 2020 Programme under grant agreement No.\ 800904. This work used the Cirrus UK National Tier-2 HPC Service at EPCC (http://www.cirrus.ac.uk) funded by the University of Edinburgh and EPSRC (EP/P020267/1)

\bibliographystyle{IEEEtran}
\bibliography{references}

\end{document}